# High-Speed Imaging of Transition from Fluid Breakup to Phase Explosion in Electric Explosion of Tungsten Wires in Air

Y. Sechrest[1], C. Campbell[2], X. Tang[2], D. Staack[2], Z. Wang[1]

[1]Los Alamos National Laboratory, Los Alamos, New Mexico 87545, USA

[2]Texas A&M University, College Station, Texas 77843-3366, USA

**Abstract:**

High-speed visible imaging of sub-microsecond electric explosion of wires at the low specific energy deposition threshold reveals three distinct modes of wire failure as capacitor charge voltage and energy deposition are increased. For 100 micron diameter gold-plated tungsten wires of 2 cm length, deposited energies of 1.9 eV/atom produces a liquid column that undergoes hydrodynamic breakup into droplets with radii of order of wire diameter on timescales of 200 microseconds. Instability growth, column breakup, and droplet coalescence follow classic Rayleigh-Plateau predictions for instability of fluid column. Above 3.2 eV/atom of deposited energy, wires are seen to abruptly transition to an expanding mixture of micron scale liquid-droplets and vapor within one frame (less than 3.33 microseconds), which has been termed 'phase explosion' in literature. Between these two limits, at 2.5 eV/atom of deposited energy, wire radius is unchanged for the first 10 microseconds before the onset of a rapid expansion and disintegration that resembles homogenous nucleation of mechanically unstable bubbles. Thermodynamic calculations are presented that separate cases by temperature obtained during heating: below boiling point, near boiling point, and exceeding boiling point.

Over its long history, the study of the electrical explosion of wires (EEW) has found many applications including: Exploding Bridge Wire (EBW) detonators; fast-pulsed, intense light sources; controlled nuclear fusion; high-power X-ray pulse generation; and nanoparticle production. The EEW process is a complex, multi-physics and materials problem with dynamics that span nano-to-micro scales in time and space, making it, even still, a challenging theoretical and observational problem. In addition, the characteristics and dynamics of the process can change dramatically with the amount and rate of deposited energy, which in turn directly impacts the use in application (e.g. efficiency in detonation, spectra and amount of produced light, or distribution of particulate products). A thorough review of the EEW process can be found in Romanova et al. (2015).

The EEW process is typically split into 'fast' and 'slow' explosions as distinguished by the rate of energy deposition, and/or electrical current density. The 'fast' explosions are characterized by the transition of the wire to a complex mixture of liquid, gaseous, and potentially even plasma phases that undergoes rapid expansions in a state that is often described as a foam-like or vapor-droplet cloud, Pikuz et al. (1999) (termed 'phase explosion'). The 'slow' explosions at low rates of energy deposition are characterized by the breakup of the wire into macroscopic pieces on longer timescales. The transition between these two modes of failure, and the change in dynamics as function of energy deposition, are of particular interest for understanding the EEW phenomenon as well as its limits in application. In this paper, we use high-speed visible imaging to investigate the transition, occurring at low specific energy deposition, from macroscopic fluid breakup of electrically heated wires to explosive phase transition. Imaging results reveal three distinct modes of wire failure at increasing deposited energies: hydrodynamic instability, rapid onset of gaseous voids, and explosive phase transition to droplet-vapor mixture.

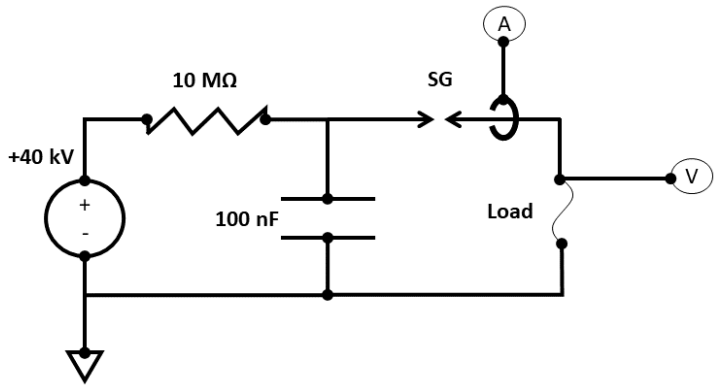

**Figure 1:** Exploding wire high-voltage pulse generator circuit diagram. A High-voltage power supply charges a 100 nF capacitor up to 40 kV. The current pulse to the load is controlled by a air-gapped, spark-gap switch (SG).

The experimental setup consists of a high voltage power supply, a single 100 nF capacitor, an air-gapped spark gap switch, and a wire load. A diagram of the circuit is shown in Figure 1. For these experiments, the capacitor is charged to between 10 and 16 kV, giving a stored energy of between 5 and 13 Joules. The high-voltage, high-current pulse to the load is controlled by an air-gapped spark gap switch which may be operated in either self-breakdown, or laser-triggered mode. The high voltage power supply is isolated from the discharge transients by a 10 megaOhm resistor in series with the capacitor. Voltage and current are measured by a high voltage probe (North Star PVM-4) and a current transformer (Pearson Model 5046), respectively. Signals were acquired by a 2 GHz oscilloscope (Teledyne LeCroy WaveRunner 204MXi, 10 GS/s) with the current signal being attenuated by a BNC attenuator (HAT-20+, 20db, 50 Ohm). The current pulse seen by the load has a peak of 1-3 kA, a rise time of 530 ns, and a characteristic frequency of 470 kHz. Inductance in the circuit is then estimated at ~1 microHenry.

The load wires were 50-100 micron diameter gold-plated tungsten wire 1-3 cm in length and 2-3% gold by weight (less than 1 micron in thickness). Estimated current densities achieved during experiments discussed herein with 100 micron gold-plated tungsten wires are between 2.5-3.2 $x\ 10^{11}\ A/m^2$. The gold coating is assumed to not have a significant impact on the thermodynamic analysis presented below, however, due to the reduced melt and evaporation temperatures of gold, the coating could play a significant role in discharge termination for cases exhibiting the phase-explosion through current shunting and other effects. These effects are not quantified.

Images were capture with a Photron SA-5 CMOS camera with 20 micron pixel size, and a camera mounted zoom lens capable of 1-4x magnification. White light backlighting was used for alignment, but videos were take of wire thermal emission with neutral density filters used to attenuate emitted light to below saturation levels. Typical camera settings were: 300,000 fps, 256x64 pixels, 369 ns global shutter speed.

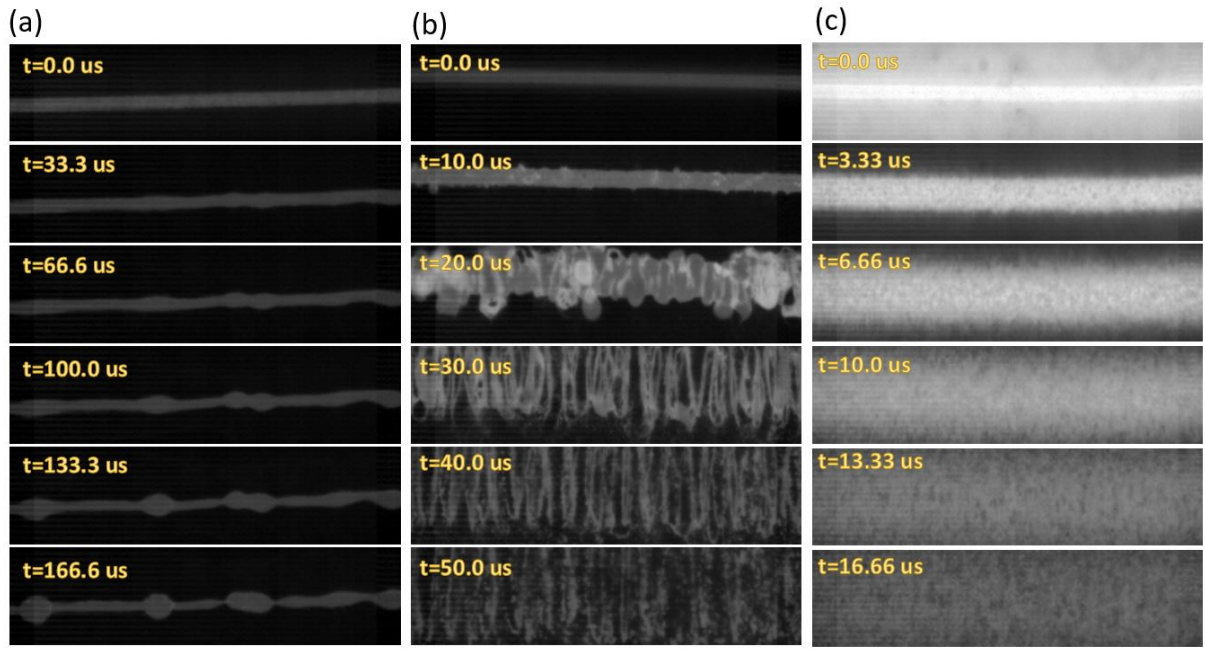

**Figure 2:** Series of frames extracted from 300k FPS high-speed video of electric explosion of gold-plated tungsten wires, 100 micron in diameter, with charge voltages of (a) 11 kV, (b) 12 kV, and (c) 13 kV. Images show (a) fluid breakup, (b) intermediate 'flash boil', and (c) phase explosion.

The transition from the fluid breakup regime to the explosive phase transition regime was probed by varying the charge voltage from 10 kV to 14 kV with 100 micron diameter gold-plated tungsten wires of 2 cm length used for the load. This effectively varies the energy deposited into the wires, and determines the maximal heating during the current pulse. Figure 2 presents a series of frames from the captured high-speed video showing the evolution of the wires for charge voltages of (a) 11 kV, (b) 12 kV, and (c) 13 kV charge voltages. For a charge voltage of 11 kV (1.9 eV/atom of deposited energy), the wire is rapidly melted by the fast current pulse, and then undergoes hydrodynamic breakup into droplets of size comparable to the wire radius on a timescale of 200 microseconds. Charge voltages above 13 kV (3.2 eV/atom of deposited energy) the wire is seen to rapidly transition into a mixture of micron scale droplets and vapor expanding outward with a velocity of approximately 30 m/s on the timescale of less than one frame (3.33 us). The 14 kV case also exhibits rapid phase explosion, and is qualitatively similar to 13 kV charge voltage. In between these extremes, at a charge voltage of 12 kV (2.5 eV/atom of deposited energy), the wire radius is largely unchanged for the first 10 microseconds, followed by an abrupt increase in wire radius and a rapid disintegration into a striated pattern of micron scale droplets over the subsequent 15 microseconds. The expansion velocity during the rapid disintegration, estimated from the change in half-max width, is 7 +/- 2 m/s. During the initial stable period, surface perturbations of scales less than the wire diameter can be seen. These small-scale perturbations give way to large perturbations on the scale of the wire diameter during the fast expansion phase that resemble rapidly expanding bubbles. At late times, these bubbles stretch in the radial direction eventually producing a pattern of voids separated by liquid ligaments that evolve into micron scale droplets. At 30 microseconds following the current pulse, there is a weak periodicity seen in cross-correlations of a 1D trace taken along the length of the wire indicating that the average scale-length for the voids is approximately 135 micron, similar to the initial wire diameter of 100 microns.

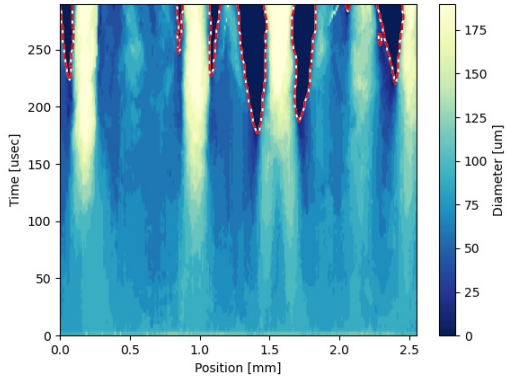

**Figure 3:** Contours of wire diameter versus time and position along length of wire. Highlighted contours at zero diameter show where detachment occurs.

The fluid breakup at low deposited energy follows classical Rayleigh-Plateau predictions for hydrodynamic instability of a fluid column. Figure 3 illustrates the time evolution of the wire diameter as a function of position along the wire. Initial perturbations have a characteristic wavenumber of $ka = 0.4 - 0.6$, where $k$ is wavenumber of the mode and $a$ is the wire radius. The classical inviscid prediction for Rayleigh-Plateau gives a must unstable mode of $ka = 0.697$, and a characteristic timescale defined by, Eggers et al. (1997):

$$t_0 = \sqrt{\frac{\rho a^3}{\gamma}} \qquad \text{Eqn 1}$$

where $\gamma = 2.5 \; N/m$ is the surface tension of liquid tungsten, and $\rho = 16800 \; kg/m^3$ is the density. This gives a characteristic time of 30 microseconds for the Rayleigh-Plateau instability of a 100 micron diameter liquid tungsten column. The observed wavelengths are slightly larger than the classical prediction, however, viscous effects are known to shift the most unstable mode to longer wavelength, Tomotika (1935). Perturbations of the wire diameter grow and coalesce into primary droplets with diameters of ~2x the initial wire diameter connected by thin ligaments. Ligaments begin to detach at 150 microseconds after the start of the current pulse, and detachment is observed to begin at the droplet surface, as demonstrated by the appearance of zero diameter contours in Figure 3 and breakage of the fluid column in Figure 2a. The detached ligaments retract, and eventually coalesces into satellite droplets.

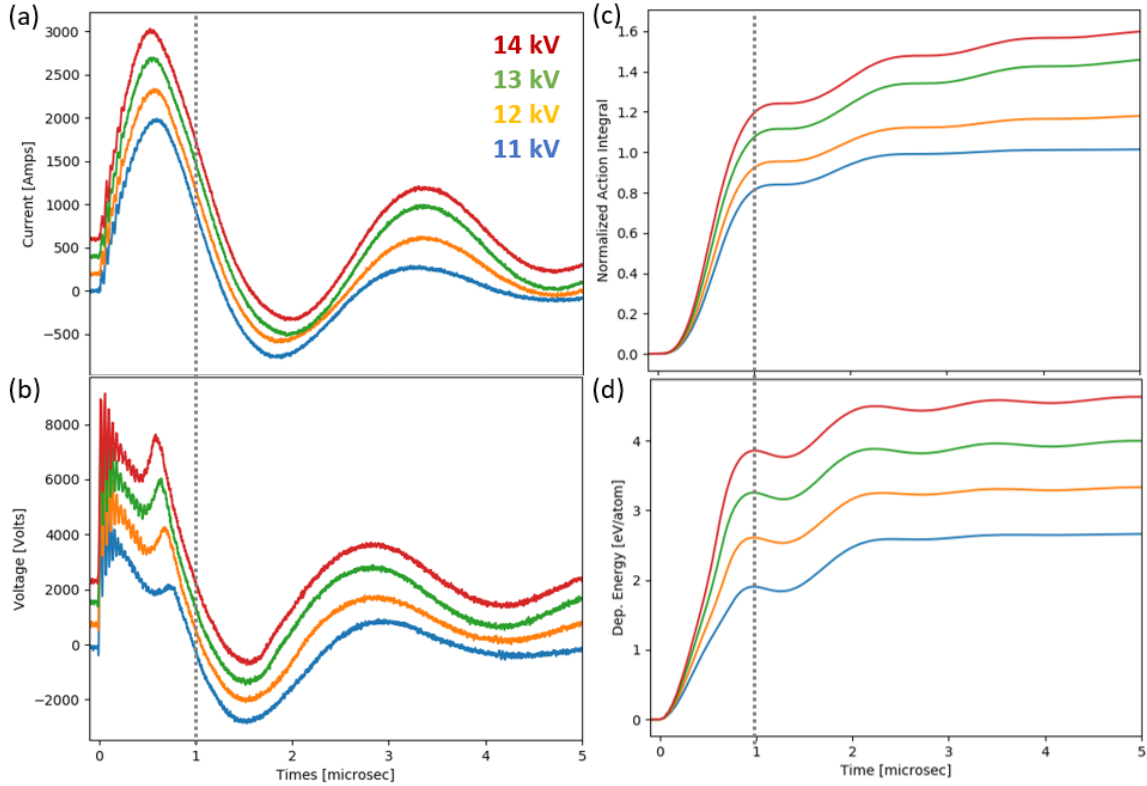

**Figure 4:** traces of (a) current, (b) voltage, (c) Normalized integral of current squared - 'Action' integral, (d) Integral of current times voltage – deposited energy. Charge voltages are 11, 12, 13, and 14 kV. Current and Voltages traces have been offset by 200 Amps and 800 Volts per trace for readability.

Current and voltage for each discharge is reported in Figure 4 (a) and (b) respectively. In addition, the normalized integral of the current squared, referred to in literature as the 'Action' integral, and the deposited energy, integral of current times voltage, are plotted in Figure 4 (c) and (d). The action integral is normalized according to the approximate equation derived in Vorob'ev et al. (1997):

$$\frac{1}{A^2 \sigma \rho C_v T_c} \int_0^\tau I^2(t)dt = T/T_c \qquad \text{Eqn 2}$$

where $A$ is the transverse cross-sectional area of the conductor, $\sigma$ is the conductivity, $\rho$ is the density, $C_v$ is the heat capacity at constant volume, $T$ is temperature, and $T_c$ is the fluid critical temperature. Values for the above quantities are taken to be for liquid Tungsten near melting point. The temperature limit for the onset of phase explosion, as determined by the nucleation mechanism described by Tkachenko et al. (2004), is $\frac{T}{T_c} = 0.9$ for the current and wire radius used in our experiments. Surprisingly, the value of $T/T_c$ estimated from measured currents and the simple approximation of equation 1, as shown in Figure 4 (c), is below the 0.9 limit after one quarter-wave of the current for the 11 kV charge voltage event that exhibits fluid breakup, and greater than the 0.9 limit for the other cases that do exhibit a rapid phase change. Still, the quantities in equation 1 are assumed to be constant with respect to temperature as the wire heats and transitions to liquid, which is a poor assumption.

To improve the estimate of $T/T_c$, we employ a 0D thermodynamic model as has been suggested by Sarkisov et al. 2007. The temperature as a function of time is found by integrating the equation

$$dT = \frac{l\,I^2(t)}{m\sigma(T)C_p(T)A} dt \qquad \text{Eqn 3}$$

Where $l$ is the length of the wire, $m$ is the mass of the wire, $C_p$ is the heat capacity at constant pressure, and other quantities are defined above. Temperature dependent conductivity (or resistivity) and heat capacity are given in Sarkisov et al. (2007). Voltage estimates from this model have been compared with experimental data of nanosecond wire explosion, and are found to be in good agreement up to the point of melting. The model does not, however, capture the voltage collapse observed in experiments, and so is expected to be inaccurate once the discharge is terminated, e.g. by current shunting around the wire core. The model also does not consider energy loss mechanisms, such as radiation, and holds the wire mass and radius fixed (i.e. no density change or wire expansion). Even with these limitations, the model gives some insight into the temperature evolution and $T/T_c$ obtained during heating. More advanced treatment requires a 1D Magneto-Hydrodymanics (MHD) model with the appropriate equation of state, as in Khishchenko et a. (2002).

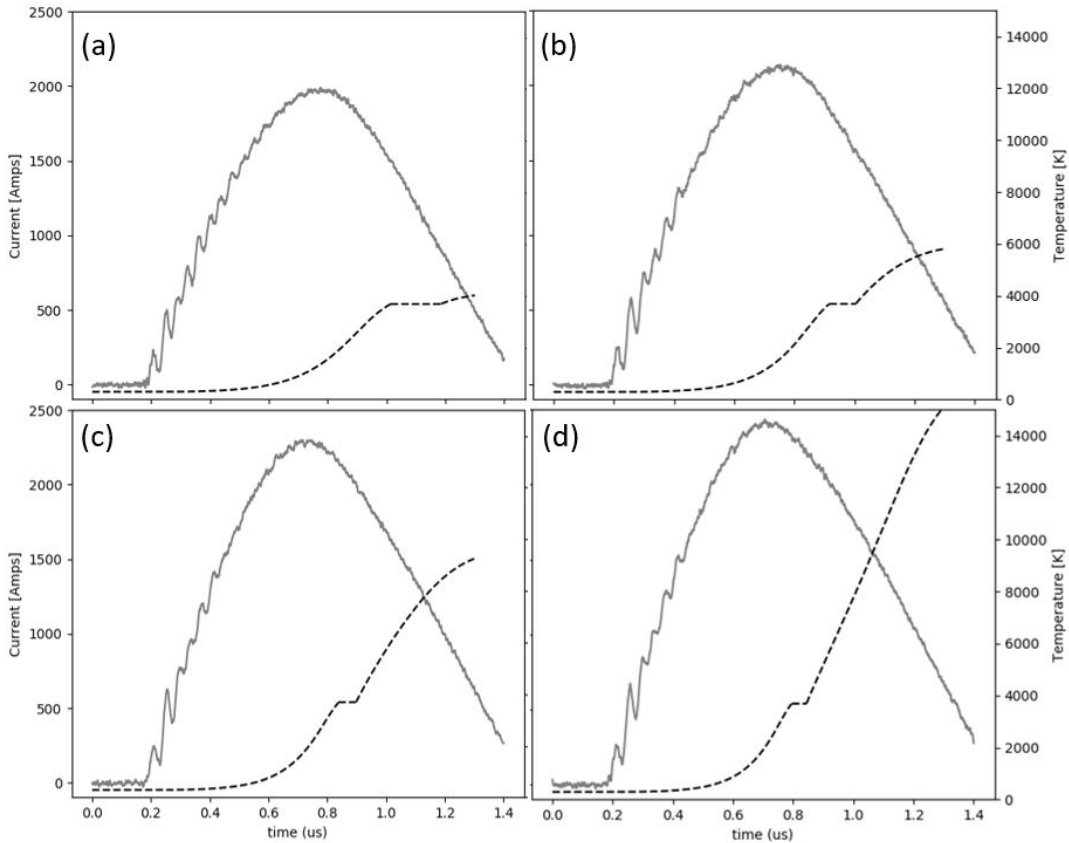

**Figure 5:** Current traces (solid) from experiment and temperature traces (dashed) calculated from Equation 2 for (a) 11 kV, (b) 12 kV, (c) 13 kV, and (d) 14 kV

Figure 5 presents the temperature time dependence as calculated using Equation 2 and the experimental current traces. For the 11 kV charge voltage, the wire is fully melted ($T_{melt} = 3695\ K$) in the first quarter-wave of the current pulse, but is not significantly heated above the melt point. At 12 kV charge voltage, the wire obtains a temperature approaching the boiling point ($T_{boil} = 6200\ K$). Higher charge voltages, greater than 13 kV, yield calculated temperature greatly exceeding the boiling point. In these discharges, MHD effects create a magnetic pressure (of order 40 MPa for the 14 kV, 2.5 kA peak current case) which impacts the liquid-gas phase transition, and result in significant overheating to temperatures above ambient boiling point, Tkachenko et al. (2004). The calculated temperatures agree qualitatively with the observed dynamics: 11 kV yields the fluid breakup of the liquid wire core on timescales much longer than the current pulse, 12 kV yields a rapid 'flash boil', and greater than 13 kV yields the phase explosion and rapid expansion of wire in mixed vapor-droplet state. The 13 kV case is, however, seen to undergo the phase explosion process, even though the calculated temperature gives $\frac{T}{T_c} \sim 0.6$ which is much below the value of 0.9 predicted for the phase explosion by nucleation mechanism.

In Summary, we report on high-speed imaging of the transition from fluid breakup to phase explosion in electrical explosion of fine wires. At low deposited energies, below 2 eV/atom, wire failure occurs via Rayleigh-Plateau instability of the liquid metal column which produces a train of droplets with diameters ~2x the diameter of the wire. At high deposited energies, greater than 3.2 eV/atom, the rapid phase explosion of the wire is observed, and the wire is seen to expand as an approximately uniform mixture of droplets and vapor. In between these extremes, a new mechanism of wire explosion is observed that is distinct from the other two cases. The intermediate explosion is characterized by an initially stable liquid metal column that undergoes rapid disintegration into a striated pattern of droplets ~10 microseconds following the current pulse. The rapid disintegration is marked by the homogenous appearance of fast expanding bubbles that give way to voids interspersed within liquid ligaments. Thermodynamic calculations indicate that the explosion mechanisms can be separated by the degree of overheating, and the temperature obtained during the current pulse. Liquid breakup is observed near the melt point, phase explosion is observed at temperatures significantly exceeding the boiling point, and the intermediate regime is observed at temperatures near the boiling point.

The authors would like to acknowledge the support of the US DOE NNSA. Los Alamos National Lab is managed by Triad National Security, LLC for the US Department of Energy's NNSA. This document has been approved for release under LA-UR-20-22711.